# SPHERICAL HARMONIC RECONSTRUCTION OF COSMIC DENSITY AND VELOCITY FIELDS

OFER LAHAV

*Institute of Astronomy, Madingley Rd., Cambridge CB3 0HA, UK*

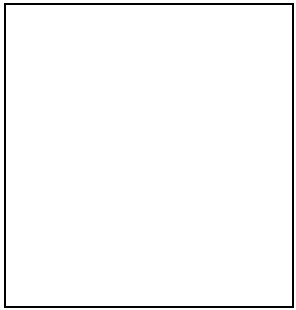

**Abstract**

The orthonormal set of Spherical Harmonics ($Y_{lm}$) provides a natural way of expanding whole-sky redshift and peculiar velocity surveys. This decomposition retains both amplitude and phase information, hence providing a unified language to describe the local cosmography as well as the power-spectrum of the galaxy density and velocity fields. We apply the method to IRAS projected and redshift surveys, to investigate redshift distortion and the value of $\Omega_0$, and to explore the Puppis cluster hidden behind the Galactic Plane. We further discuss the choice of radial function and the estimation of the coefficients in 3-D, and the removal of shot-noise and correction for incomplete sky-coverage by a 'Wiener filter'. The data compressed by the Harmonic decomposition can then be analysed by Artificial Neural Networks in comparison with simulations.

## 1 Introduction - why Spherical Harmonics ?

The rapid progress made in observations of large scale structure and in particular the production of 'nearly all-sky' redshift surveys, calls for new approaches to quantify the galaxy distribution and peculiar motions. The much-discussed statistics such as correlation functions and 'counts-in-cells' are most useful for estimating the underlying power-spectrum of the density field, but ignore the phase information, which is so crucial for describing features such as the Supergalactic Plane.

In the light of new IRAS and optical 'whole-sky' redshift surveys and the intensive work on peculiar velocity surveys it is natural to discuss the density and velocity fields by means of Spherical Harmonics. In brief, the density field is expanded using the orthogonal set of Harmonics, $Y_{lm}(\theta, \phi) \propto P_l^{|m|}(\cos\theta) \exp(im\phi)$, where $\theta$ and $\phi$ are the spherical polar angles, and $P_l^{|m|}$'s are the associated Legendre Polynomials of degree $l$ and order $m$. Generally a 3-D density field can be expanded out to $l_{max}$ by

$$\rho(\mathbf{r}) = \sum_l \sum_{m=-l}^{+l} a_{lm}(r) Y_{lm}(\hat{\mathbf{r}}), \qquad (1)$$

where the radial coefficients $a_{lm}(r)$ are estimated using an appropriate choice of weighting function $f(r)$ and the orthogonality property of the Harmonics by

$$a_{lm}(r) = \sum_{gal} f(r_i) Y_{lm}^*(\hat{\mathbf{r}}_i) , \qquad (2)$$

where the sum is over the observed galaxies. Spherical Harmonic Analysis (SHA) has been discussed for analysing projected surveys about 20 years ago [19] but was not that useful given the poor sky coverage of the samples existing at the time. More recently SHA has been reconsidered and applied to IRAS surveys ([3], [24], [25], [13], [6], [23] ) and to the peculiar velocity field ([21], [15], [11]).

Some of the advantages of SHA in probing the large scale structure are:

(i) The Harmonics retain both amplitude and phase information, hence providing a unified language for both cosmography and statistics.

(ii) A spherical orthogonal coordinate system is the natural one for analysing whole-sky distributions.

(iii) The SHA provides natural smoothing with angular resolution $\sim \pi/l$. It covers a wide range of scales, from the dipole to the small scales usually probed by correlation functions.

(iv) The SHA is an efficient data compression procedure. The compressed data can then be used as input to various statistical tools such as Maximum Likelihood, reconstruction methods, or Artificial Neural Networks (see below).

(v) The estimated coefficients can be used for comparison of different tracers of the density field (e.g. IRAS and optical), as well as with the peculiar velocity velocity field and Background radiations.

(vi) The much-discussed gravity dipole ($l = 1$) and quadrupole ($l = 2$) are only special cases of the SHA.

(vii) At small angles the SHA becomes the familiar 2-D Fourier transform. The mean-square of amplitudes is simply related to the 3-D power-spectrum or the 2-point correlation function.

One might argue that SHA assumes that the observer is at the centre of the universe and is therefore a biased statistic. However, shot-noise increases with radial distance from us in flux limited surveys. Moreover, the analysis can be done from other centres as well. Another major problem of the SHA is incomplete sky coverage, in particular due to obscuration and/or confusion by the Galactic Plane. This can be accounted for by modelling an angular window function([19], [24]), or by a Wiener reconstruction ([14] and below).

Here we discuss several examples which illustrate the utility of the Harmonics in overcoming some of the key problems in the study of the large scale structure: shot-noise, redshift distortion and incomplete sky coverage, and in estimating properties such as the power-spectrum, biasing and the density parameter $\Omega_0$. We summarize studies on a projected IRAS sample of galaxies brighter than 0.7 Jy [24], the 2Jy IRAS redshift survey [25], and the Puppis cluster of galaxies behind the Galactic Plane [13]. We then review the Harmonics as probes of the power-spectrum and cosmological parameters, in particular using the SHA to analyse redshift distortion [6] and to estimate the combination of the density and bias parameters $\Omega_0^{0.6}/b$. Then we discuss on-going projects of Wiener filter reconstruction of noisy surveys with incomplete sky coverage [14], a 3-D orthogonal expansion and pattern recognition with Harmonics and Artificial Neural Networks. Further details and discussion on Harmonics are given in this volume by K. Fisher (on redshift distortion) and by C. Scharf (on cross-correlation of optical and IRAS Harmonics).

## 2   Cosmography with Spherical Harmonics: the Puppis cluster

The expansion in Spherical Harmonics can be used to reconstruct the surface density or brightness up to a certain Harmonic $l_{max}$. Various reconstructions by Scharf et al. ([24], [25] [13] and [23]), reveal familiar and new structures, and illustrate for example the 'tug of war' between the Great Attractor/Centaurus and Perseus-Pisces superclusters. Figure 1 in ref. [24] indicated a significant overdensity in the direction of Puppis ($l \sim 240^o; b \sim 0^o$). This cluster has also been noticed and studied independently by other groups ([10] and [28]). To quantify the importance of Puppis relative to other structures in the Local Universe we supplemented the IRAS 2 Jy $|b| > 5^o$ redshift survey [27] with an IRAS-selected sample in the direction of Puppis ($|b| < 5^o; 230^o < l < 260^o$), which consists of 32 identified galaxies, 12 of them with measured redshift. We find [13] that the projected number counts of galaxies brighter than 2 Jy in Puppis is about half that of Virgo. A Spherical Harmonic reconstruction (Figure 1) shows that out to a distance of 3000 km/sec Puppis is second only to Virgo. We estimate that Puppis (which lies below the Supergalactic Plane) may contribute at least 30 km/sec to the motion of the Local Group perpendicular to the Supergalactic Plane ($V_{SGZ} = -370$ km/sec in the Cosmic Microwave Background frame). Together with the Local Void (above the Supergalactic Plane) and Fornax and Eridanus (below it) this may explain the origin of the so-called "Local Anomaly".

Figure 1: Spherical Harmonic reconstruction with coefficients up to $l_{max} = 10$, of galaxies with 500 km/sec $< cz_{LG} <$ 3000 km/sec in the 2Jy IRAS ($|b| > 5^o$) combined with our Puppis sample ($|b| < 5^o$). Plots are equal area hemispheres with the left-hand side plot centred on Galactic $l = 240°$, $b = 0°$. The Galactic Plane runs horizontally across the plots (solid line), dashed lines bound the region $|b| < 5°$ and longitudes are indicated. The South Galactic hemisphere is at the top. The lightest solid contour is at the mean, dashed and solid contours indicate densities below and above the mean respectively. Contour separation is 3 times the shot-noise level. Associations with local structures are labelled: V - Virgo, PU - Puppis, F - Fornax E - Eridanus, Leo - Leo, Cen - Centaurus, PIT - Pavo-Indus-Telescopium, Cam - Camelopardalis and UM - Ursa Major. From ref. [13].

## 3 Spherical Harmonics as probes of the power-spectrum

The mean-square of Harmonics can be related to the power-spectrum of mass fluctuations $P_m(k) = \langle |\delta_k|^2 \rangle$ in Fourier space for a variety of cosmic phenomena. A useful identity in deriving the relations below is the expansion of a plane wave in spherical waves, $e^{i\mathbf{k}\cdot\mathbf{r}} = 4\pi \sum_l \sum_m (i)^l j_l(kr) Y_{lm}(\hat{\mathbf{r}}) Y_{l-m}(\hat{\mathbf{k}})$, where $j_l$ is the spherical Bessel function.

### 3.1 Galaxy Harmonics in real space

If the distances to galaxies are known then an estimator for the Harmonic coefficients is $a_{lm}^R = \sum_{gal} f(r_i) Y_{lm}^*(\hat{\mathbf{r}_i})$, where $f(r)$ is a 'weighting function'. The mean-square prediction is ([24], [25]):

$$\langle |a^R_{lm}|^2 \rangle = \frac{2}{\pi} b^2 \int dk\, k^2 P_m(k) \left| \Psi^R_l(k) \right|^2 , \qquad (3)$$

where the 'window function' which depends on the radial selection function $\phi(r)$ is

$$\Psi^R_l(k) = \int dr\, r^2 \phi(r) f(r) j_l(kr) . \qquad (4)$$

We have assumed that the power-spectrum of the galaxies is $b^2 P_m(k)$ ('linear biasing'). To this 'cosmic scatter' one should also add the 'shot noise' due to the discreteness of objects, $\langle |a_{lm}|^2 \rangle_{sn} = \int dr\, r^2 \phi(r) f^2(r) \sim \frac{1}{4\pi} \sum f_i^2$.

### 3.2 Galaxy Harmonics in redshift space

The power-spectrum derived from a redshift survey will differ from the one in real-space (Kaiser[9]). This is also seen in the Harmonic analysis of redshift surveys (Scharf & Lahav[25]), and the Harmonics in redshift space can be formulated (Fisher, Scharf & Lahav[6]) for a flux-limited survey in linear theory as:

$$\langle |a^S_{lm}|^2 \rangle = \frac{2}{\pi} b^2 \int dk\, k^2 P_m(k) \left| \Psi^R_l(k) + \frac{\Omega^{0.6}_0}{b} \Psi^C_l(k) \right|^2 , \qquad (5)$$

where the redshift distortion window function is

$$\Psi^C_l(k) = \frac{1}{k} \int dr\, r^2 \phi(r) \frac{df}{dr} \left[ j'_l(kr) - \frac{1}{3}\delta_{l1} \right] . \qquad (6)$$

For $l=1$ (dipole) this recovers Kaiser's 'rocket effect' [9]. In the special case that both $\phi(r)$ and $f(r)$ are power laws, the real and redshift Harmonics are simply related by $\langle |a^S_{lm}|^2 \rangle = (1 + \beta T_l)^2 \langle |a^R_{lm}|^2 \rangle$, where $T_l$ depends on the power-indices of the selection and weighting functions, and $\beta \equiv \Omega^{0.6}_0/b$.

We have applied (ref. [6] and K. Fisher in this volume) eq. (5) and a Maximum Likelihood analysis to the 1.2 Jy redshift survey ([4]). If we fix the normalization of IRAS galaxies in real space ($\sigma_8 = 0.69 \pm 0.04$) and solve for the shape parameter ($\Gamma \equiv \Omega h$) of a CDM power-spectrum, we get $\Gamma = 0.17 \pm 0.05$ and $\beta = 0.94 \pm 0.17$ ($1-\sigma$). This value of the shape-parameter is in accord with other measurements of excess power on large scale (e.g. [5]), and $\beta$ for IRAS (which is strongly coupled with $\sigma_8$) is again found to be higher than $\beta$ deduced from optical studies (e.g., [16], [12], and M. Hudson in this volume). It is interesting to note that although we have not used here any independent measurement of peculiar velocities (i.e. distance indicators), our result for $\beta$ is in agreement, and of comparable accuracy, to the results deduced by comparing the peculiar velocity field (POTENT) with IRAS redshift surveys (e.g. [2]).

### 3.3 Peculiar Velocity Harmonics

The Harmonics for the peculiar velocities are simply the difference between Harmonics in redshift space and real space:

$$\langle |a^S_{lm} - a^S_{lm}|^2 \rangle \propto \langle |V_{lm}|^2 \rangle = \Omega^{1.2}_\circ H^2_\circ \frac{2}{\pi} \int dk\, k^2 P_m(k) \left| \Psi^C_l(k) \right|^2 , \qquad (7)$$

but in the case of incomplete sky coverage the window function is more complicated [21].

### 3.4 The Microwave Background (Sachs-Wolfe)

Finally, the much-discussed Sachs-Wolfe fluctuations in the Microwave-Background are written in terms of Harmonics as:

$$\langle |a_{lm}|^2 \rangle_{SW} = \left( \frac{H_\circ}{2c} \right)^4 \frac{2}{\pi} \int \frac{dk}{k^2} P_m(k) \left[ j_l(2ck/H_\circ) \right]^2 \qquad (8)$$

# 4 Wiener filter for Spherical Harmonic reconstruction

The analysis of whole-sky galaxy surveys commonly suffers from the problems of shot-noise (due to the discreteness of objects) and incomplete sky coverage (e.g., at the Zone of Avoidance). Here we discuss a method [14] for correcting for these effects using a Bayesian framework, and the orthogonality property of the Harmonics. The recovery of complete distribution from noisy and incomplete data is a classic problem of inversion. It is well known that a straightforward inversion is unstable, and hence a regularisation of some sort is essential. In the Bayesian spirit we are using here raw data and a 'prior model' to produce 'improved data'.

We formulate our problem as follows: What are the full-sky noise-free Harmonics $a_{lm}$ given the observed Harmonics $c_{lm,obs}$, the mask $W$, and a prior model for the power-spectrum of fluctuations? The observed Harmonics (with the masked regions filled in uniformly according to the mean) are related to the underlying 'true' whole-sky Harmonics by (cf. [20], eq. 46.33)

$$c_{lm,obs} = \sum_{l'} \sum_{m'} W_{ll'}^{mm'} [a_{l'm'} + \sigma_a] \qquad (9)$$

where the monopole term ($l' = 0$) is excluded, and we have added the shot-noise for number-weighted $a_{lm}$ Harmonics with variance $\langle \sigma_a^2 \rangle = \mathcal{N}$ (the mean number of galaxies per steradian, independent of $l$ in this case).

By the rule of conditional probability and the assumption that the density fluctuations are drawn from a Gaussian random field we can write

$$P(\mathbf{a}|\mathbf{c}_{obs}) = \frac{P_G(\mathbf{a}, \mathbf{c}_{obs})}{P_G(\mathbf{c}_{obs})} \ , \qquad (10)$$

where the vector $\mathbf{c}_{obs}$ represents the set of observed Harmonics $\{c_{lm,obs}\}$ and $P_G$ stands for a Gaussian distribution function with variance and covariance which depend on an assumed power-spectrum. This is a special case of constrained realizations formalism ([8] and Y. Hoffman in this volume), but here the formulation and computation are greatly simplified due to the orthogonality of the Harmonics. The full derivation of maximizing the probability (eq. 10) with respect to $\mathbf{a}$ will be given elsewhere (Zaroubi et al., in preparation), but the answer for the 'mean field' reconstruction is simply

$$\hat{\mathbf{a}} = \mathbf{F} \mathbf{W}^{-1} \mathbf{c}_{obs} \ , \qquad (11)$$

with

$$\mathbf{F} = diag \left\{ \frac{\langle a_l^2 \rangle_{th}}{\langle a_l^2 \rangle_{th} + \langle \sigma_a^2 \rangle} \right\} . \qquad (12)$$

One can also express the scatter around this value. It can be shown that this result also gives the minimum of the variance $\langle |\hat{\mathbf{a}} - \mathbf{a}|^2 \rangle$ for a desired filter matrix $\mathbf{F}$. This $\mathbf{F}$ matrix is in fact the well-known Wiener filter (the ratio of signal to signal+noise) commonly used in signal processing (e.g. [22]). Note that it requires *a priori* knowledge of the variances in both the signal and the noise. When the noise is negligible the factor is approaching unity, but when it is significant the measurement is attenuated accordingly. Even if the sky coverage is $4\pi$ ($\mathbf{W} = \mathbf{I}$), the Wiener filter can be used to remove shot-noise, giving the most probable picture of the underlying 'continuous' density field. In the case of full sky coverage, only the amplitudes are affected by the correction, not the relative phases. For example, the dipole direction is not affected by the shot-noise, only its amplitude. Of course, if the sky coverage is incomplete, both the amplitudes and the phases are corrected. The reconstruction also depends on how many Harmonics are observed ($l_{max}$) and how many are desired for the reconstruction ($l'_{max}$). Note also that the method is *non*-iterative.

Here we apply the method to the 2-D sample of 5313 IRAS galaxies brighter than 1.2 Jy ([4], [5] ), covering 88 % of the sky. Practically, rather than inverting $\mathbf{W}$ (eq. 11) we solve the equation $\mathbf{c_{obs}} = \mathbf{W}\mathbf{F}^{-1}\hat{\mathbf{a}}$ using the Singular Value Decomposition algorithm. The 'best reconstruction' is shown in Figure 2, indicating the connectivity of the Supergalactic Plane where crossed by the Galactic Plane, and confirming the Puppis cluster.

Figure 2: Wiener reconstruction of the 2-D 1.2 Jy IRAS galaxy sample, for Harmonics $l \leq 15$, plotted in Aitoff Galactic projection. The reconstruction corrects for incomplete sky coverage, as well as for the shot-noise. The assumed prior model is a low density CDM (with shape parameter $\Gamma = 0.2$ and normalization $\sigma_8 = 0.7$), although a prior of standard CDM gives a very similar reconstruction. The reconstruction indicates that the Supergalactic Plane is connected across the Galactic Plane at Galactic longitude $l \sim 135°$ and $l \sim 315°$. The Puppis cluster stands out at the Galactic Plane at $l \sim 240°$. The horizontal dashed lines at $b = \pm 5°$ mark the major 'Zone of Avoidance' in the IRAS sample.

Preliminary application to simulated samples with larger incompleteness indicate bias in the method, which we are currently exploring. The method can be extended to 3-D in both real and redshift spaces, and applied to other cosmic phenomena such as the COBE Microwave Background maps.

## 5   3-D orthogonal Harmonic expansion

An obvious extension of the above methods is to three dimensions. We expand the fluctuations in the density field in Spherical Harmonics $Y_{lm}$ and Spherical Bessel functions $j_l(z)$ (cf. ref. [1]):

$$\rho(\mathbf{r}) - \bar\rho = \sum_l \sum_m \rho_{lm}(r) Y_{lm}(\hat{\mathbf{r}}) = \sum_l \sum_m \sum_n c_{nl}\, \rho_{lmn}\, j_l(k_{nl'}\, r)\, Y_{lm}(\hat{\mathbf{r}}) , \qquad (13)$$

for $l > 0$. Similarly, the fluctuation in the potential can be expanded as $\Psi = \sum_l \sum_m \Psi_{lm}(r) Y_{lm}(\hat{\mathbf{r}})$, and the two quantities are related by Poisson equation, $\nabla^2[\Psi_{lm}(r)Y_{lm}(\hat{\mathbf{r}})] = -4\pi G \rho_{lm}(r) Y_{lm}(\hat{\mathbf{r}})$. Let us assume that the data are given within a sphere of radius $a$, such that inside the sphere the desired density fluctuation is specified by $\rho_{lm}(r)$, but for $r > a$ the fluctuation is $\rho_{lm}(r) = 0$ (this simply reflects our ignorance about the density field out there; the fluctuations do not vanish of course at large distances). Hence, inside the sphere $\rho_{lm}(r) \propto k^2 \Psi_{lm}(r) \propto j_l(kr)$, and outside the sphere $\Psi_{lm}(r) \propto r^{-(l+1)}$. It is sensible to require the potential and its logarithmic derivative to be continuous at $r = a$ (D. Lynden-Bell, private communication). It then follows that the discrete $k$'s are the zeros of $j_{l-1}(k_{nl'}a) = 0$, where $l' = l - 1$. It can be shown that this condition ensures the orthogonality of the

Figure 3: Reconstruction of 3-D Harmonic expansion (in redshift space) with $\rho_{lmn}$ coefficients up to Harmonic $l_{max} = 15$, evaluated at $r = 1000$ km/sec. The Local Supercluster appears here centred at Galactic coordinates $(l \sim 290°; b \sim 45°)$ and the Local Void at $(l \sim 45°; b \sim 0°)$.

Bessel functions within the sphere, $\int_0^a dr \, r^2 j_l(k_{nl}r) \, j_l(k_{n'l}r) = \delta_{nn'} c_{nl}^{-1}$ with $c_{nl}^{-1} = \frac{1}{2}a^3(k_{nl}a)j_l^2(k_{nl}a)$, where $c_{nl}$ is also the normalization in eq. (13). Note that the Bessel function $j_l(z)$ has an infinite number of zeros, which behave asymptotically like $(k_{nl}a) \sim \pi(n+l/2)$. We therefore need to choose the maximal number of radial modes $n_{max}(l)$ for a desired spatial resolution. Other boundary conditions are also possible, although they are somewhat less natural, e.g. one can choose (ref. [1]) the $\rho_{lm}(r)$ to vanish at $r = a$, corresponding to the zeros of $j_l(k_{nl}a) = 0$. One can also choose a different set of radial functions (e.g. the associated Laguerre polynomials [15]), but the spherical Bessel functions naturally appear in our problem (see Section 4).

The above formalism only holds for a volume-limited sample. In practice, given a flux-limited sample we can estimate the coefficients from the data by

$$\hat{\rho}_{lmn} = \sum_{gal} \frac{1}{\phi(r_i)} \, j_l(k_{nl'} \, r_i) \, Y_{lm}^*(\hat{\mathbf{r}}_i) \,, \tag{14}$$

where the sum is over galaxies with $r < a$, and $\phi(r)$ is the selection function.

As an example of the method, we have applied it to the 2 Jy IRAS redshift survey [27] with $a = 6000$ km/sec, including 1888 galaxies in that sphere. The number of radial modes $n_{max}(l)$ was chosen to give a desired resolution of $k_{nl'} < 21/a$ and $l_{max} = 15$, resulting in 617 $\rho_{lmn}$ coefficients. We show a reconstruction (in redshift space) by evaluating $\hat{\rho}_{lm}(r) = \sum_n c_{nl}\rho_{lmn}j_l(k_{nl'}r)$ at $r = 1000$ km/sec and reconstructing the angular Harmonic expansion up to $l_{max} = 15$. Figure 3, in Galactic coordinates, shows the Local Supercluster and the Local Void in full glory. However, at larger distances the reconstruction is getting more noisy, in part because of the $1/\phi(r)$ weighting scheme. We are currently exploring other weighting schemes and the incorporation of a 'Wiener filter' to improve the quality of the 3-D reconstruction at larger distances.

## 6 Observed vs. simulated Harmonics using Artificial Neural Networks

The comparison of the observed large-scale structure with cosmological simulations remains a challenge. The traditional methods, like the 2-point correlation function, and counts-in-cells, do not represent the *phase* information. In simple words, two realisations which have very different apparent features could nevertheless have the same correlation functions. Another point concerns the methodology of comparing models with observations. Usually, by performing a $\chi^2$ or maximum likelihood test we estimate the probability for the data given a model, $P(data|model)$. But what is really of interest is the probability for the model given the data, which is given by Bayes' theorem,

$$P(model|data) \propto P(model) \, P(data|model). \tag{15}$$

Artificial Neural Networks (ANN) incorporate very well the aspects of phase information and the Bayesian approach, without ad-hoc assumptions, e.g. that the Harmonics are drawn from a Gaussian density field. ANN algorithms were originally developed to model the human brain. Here we use an ANN model known as the Backpropagation algorithm (e.g. [7]). It consists of nodes (analogous to human neurons) arranged in a series of layers. In our case the input layer consists of the 24 Harmonic coefficients, and the output layer consists of the 3 models ('classes'). We also include 'hidden layers' which allow non-linearity in a complex classification space. ANNs have recently been applied to several problems in Astronomy, e.g. morphological classification of galaxies [26]. Non-astronomical applications somewhat similar to our problem are face and speech recognition and identification of hand-written characters. The ANN 'learns' by minimizing least-squares and fixing its free parameters, known as the 'weights', from a sample for which the answer is known (simulations in our case). Then it can predict classification for new data. The output vector can be viewed as the Bayesian *a posteriori* probability for a model given the data. Moreover, the sum of the output vector components is nearly unity, as expected for a probabilistic classifier. Here we train ANN on simulated Harmonics for which the assumed model is known, and then present the ANN with the observed Harmonics, and ask it to choose a favourite model, a question usually addressed in a limited way by statistics such as the 2-point correlation function, or by visual impression.

As an example, the observed projected (number-weighted) distribution of IRAS galaxies brighter than 0.7 Jy [17] was expanded in Harmonics $1 \leq l \leq 4$, i.e. the galaxy distribution was compressed into 24 coefficients. The simulations [18] are of standard biased Cold Dark Matter (CDM) model, unbiased scale-free power-spectrum, $k^n$ with $n = -1$ (which has more power on large scales than the standard CDM model) and a Poisson model. All simulations mimic the observed IRAS selection effects. We have used 24 simulations of each model, i.e. 72 simulations in total (i.e. assigning equal prior probability of a $\frac{1}{3}$ for having each of the models). The Network configuration is (24;2;3), i.e. it includes 24 input Harmonic coefficients, 2 hidden units, and 3 output nodes representing the models.

If presented with randomly oriented data the Net will be confused, since the individual $a_l^m$'s are dependent on the coordinate system (only the mean-square for each $l$ is invariant under rotation). To to measure *relative phases*, we first aligned the dipole and quadrupole of each simulation with that of the IRAS sample. We then carried out (O.L. & C. Scharf) the following experiments:

(1) The ANN was trained on all 72 realisations, and was then given the same 72 realisations to classify. The success rate was 94%, i.e. the Net 'memorized' very well all the cases it has seen. While this may seem very impressive at first sight, it also indicates a possible problem. When the number of weights (59 in this case) is large relative to the number of data points ($72 \times 24 = 1728$), we might over-fit the data.

(2) As a more challenging task for the Net, we divided the 72 simulations into two, a training set of 48 simulations (16 for each model), and a testing set of 24 simulations (8 for each model). After training the net on the 48 simulations we presented it with the 24 simulations it had never seen before. the Net recognised all 8 Poisson simulations correctly, as the human eye will easily do. Of the 8 CDM realisations it recognized 7 correctly, and 1 wrongly as Poisson. Of the $n = -1$ realisations it classified only 3 realisations correctly, and 5 wrongly as CDM.

(3) The Net, trained on the 72 realisations as in (1), was then shown the single 'true' IRAS observation. To our surprise the Net assigned 100 % probability of belonging to a CDM model (and 0 % for the other two models). It was puzzling that the Net favoured one model so strongly. But then we realized that one of the 72 realisations happened to have Harmonic coefficients very similar to those of IRAS. On one hand it indicates the strength of the Net in recognizing similar objects. But it raises the question of 'fair statistics', i.e. how common such a simulation is.

(4) To cope with the problem of the small number of simulations we generated 10 Bootstrap combinations of realisations (allowing repetition). We trained the Net on each of them and presented each trained Net with the IRAS Harmonics. In some cases the Net favoured a CDM model, in other cases it favoured $n = -1$ model, but rarely the Poisson model. The average (with large scatter) is 27% 45% and 17% for CDM, $n = -1$ and the Poisson models, respectively. The fact that the probabilities do not sum up to 100% indicates that our Net is not yet perfect, due to the small number of simulations.

These results, hinting at excess power relative to standard CDM, are in accord with more conventional statistical method of Maximum Likelihood for the Harmonic amplitudes ([24], [25]) and other studies (e.g. [5]) To explore to what extent the Net actually used the phase information we also calculated the cross-correlation $\langle a_{l,IRAS}^m \, a_{l,sim}^m \rangle$ and found, not too surprisingly, that for $l \leq 4$ (large scales) they are consistent with random phases (which were the initial conditions of the simulations). Hence in this example it is likely that the Net mainly made inference on the basis of the amplitudes, rather than the phases. It would be interesting to extend the analysis to the non-linear scales (high $l$'s) for which the phases are non-random, and to simulations in which the initial conditions had non-random phases, e.g. cosmic strings and texture. An improvement to the statistic can be achieved by presenting the Net with much larger number of independent simulations.

While the conventional approach of comparing rms values of the Harmonics or correlation functions is more straightforward and can be used to reject models, our approach which incorporates phases is useful when it is difficult to discriminate between models based on the amplitudes alone. As with any statistic, the result depends on what physical information is fed into the Net. The method could be generalized to present the Net with data from redshift surveys and with peculiar velocities of galaxies.

## 7 Discussion

We have presented several examples of application of Spherical Harmonics Analysis to the study of the large scale structure, for both cosmographical (e.g., the Puppis cluster) and cosmological inference (e.g., the power-spectrum and $\Omega_0^{0.6}/b$). We are currently developing the Harmonic expansion in 3-D further, correcting simultaneously (using the properties of the Harmonics) for incomplete sky-coverage, redshift distortion and the shot-noise by Wiener filter (Fisher, Hoffman, O.L., Lynden-Bell & Zaroubi, in preparation). This procedure will then be applied to new all-sky IRAS and optical redshift surveys, and to surveys of the peculiar velocity field. The 3-D $\rho_{lmn}$ coefficients will allow objective non-parametric comparison of different surveys of light and mass in the local universe.

**Acknowledgements.** I am grateful to K. Fisher, Y. Hoffman, D. Lynden-Bell, C. Scharf and S. Zaroubi for their contribution to the work presented here and for many stimulating discussions.